\documentclass{nature}
\usepackage{graphicx}
\usepackage{dcolumn}
\usepackage{bm}
\usepackage{amsmath}
\usepackage{amssymb}
\usepackage{color}
\usepackage{tikz}
\usepackage{caption}
\usepackage{lineno}
\title{Quantum Simulation of the Non-Fermi-Liquid State of Sachdev-Ye-Kitaev Model}

\author{Zhihuang Luo$^{1,2,3,4}$, Yi-Zhuang You$^{5}$, Jun Li$^{3,6,7}$, Chao-Ming Jian$^{8,9}$, Dawei Lu$^{3,6,7\ast}$, Cenke Xu$^{10}$, Bei Zeng$^{3,4,11\ast}$, $\&$ Raymond Laflamme$^{4,12}$}

\begin{document}

\maketitle

\begin{affiliations}
 \item Beijing Computational Science Research Center, Beijing, 100193, China.
 \item Laboratory of Quantum Engineering and Quantum Metrology, School of Physics and Astronomy, Sun Yat-Sen University (Zhuhai Campus), Zhuhai 519082, China
 \item Shenzhen Institute for Quantum Science and Engineering, and Department of Physics, Southern University of Science and Technology, Shenzhen 518055, China.
 \item Institute for Quantum Computing and Department of Physics and Astronomy, University of Waterloo, Waterloo N2L 3G1, Ontario, Canada.
\item  Department of Physics, Harvard University, Cambridge, MA 02138, USA.
\item Center for Quantum Computing, Peng Cheng Laboratory, Shenzhen 518055, China
\item Shenzhen Key Laboratory of Quantum Science and Engineering, Shenzhen 518055, China
\item Station Q, Microsoft Research Santa Barbara, California 93106, USA.
 \item Kavli Institute of Theoretical Physics, University of California, Santa Barbara, California 93106, USA.
 \item Department of Physics, University of California, Santa Barbara, California 93106, USA.
 \item Department of Mathematics and Statistics, University of Guelph, Guelph N1G 2W1, Ontario, Canada.
 \item Perimeter Institute for Theoretical Physics, Waterloo N2L 2Y5, Ontario, Canada.
\end{affiliations}

\begin{abstract}
The Sachdev-Ye-Kitaev (SYK) model incorporates rich physics,
ranging from exotic non-Fermi liquid states without quasiparticle
excitations, to holographic duality and quantum chaos. However,
its experimental realization remains a daunting challenge due to
various unnatural ingredients of the SYK Hamiltonian such as its
strong randomness and fully nonlocal fermion interaction. At
present, constructing such a nonlocal Hamiltonian and exploring
its dynamics is best through digital quantum simulation, where
state-of-the-art techniques can already handle a moderate number
of qubits. Here we demonstrate a first step towards simulation of
the SYK model on a nuclear-spin-chain  simulator. We observed the fermion paring instability of
the non-Fermi liquid state and the chaotic-nonchaotic transition
at simulated temperatures, as was predicted by previous theories.
As the realization of the SYK model in practice, our
experiment opens a new avenue towards investigating the key
features of non-Fermi liquid states, as well as the quantum
chaotic systems and the AdS/CFT duality.
\end{abstract}

\section*{Introduction}

The Sachdev-Ye-Kitaev (SYK) model describes a strongly interacting
quantum system with random all-to-all couplings among $N$ Majorana
fermions~\cite{SachdevYe1993,Kitaev2015,Sachdev2015,Polchinski2016,MaldacenaStanford2016,Witten2016,Klebanov2016,Gross2016,Fu:2016it,Krishnan2017,You:2017gd,Banerjee:2017sf,Fu2017,Gu2017,Gu:2017zh,Chen2017,Murugan2017,Narayan2017,Chew:2017jt,Jian:2017rr,Peng:2017vl,Yoon2017,Chen:2017yg,Jian:2017db,Zhang:2017cs,Narayan:2017zp}.
At large $N$, this model is exactly solvable and exhibits an
explicit non-Fermi liquid (NFL) behavior with nonzero entropy
density at vanishing temperature. In condensed matter physics, the
most well-known (yet poorly understood) NFL is the ``strange
metal" phase at optimal doping of the cuprates high temperature
superconductors, where the resistivity scales linearly with
temperature for a very large range in the phase
diagram~\cite{linear1,linear2,linear3,Varma1989}, as shown in Fig.
\ref{fig:phase}. The strange metal phase can be viewed as the
parent state of high $T_c$ superconductors, in contrast to the
role of Fermi liquid in ordinary BCS superconductors. Very
recently a number of works have constructed non-fermi liquid
states based on SYK physics, with potential applications to
condensed matter
systems~\cite{Song2017,sachdev2018,senthil2018,strangemetal}. The
SYK model and its generalized SYK$_q$ models with a $q-$fermion
interaction also attract tremendous interests in the quantum
information and string theory community. For example, the maximal
chaotic behavior for $q
> 2$ grant the model a holographic dual to the $(1+1)d$ Einstein
gravity with a bulk black
hole~\cite{Sachdev2010,Sachdev2015,Polchinski2016,jensen2016,verlinde2016,MaldacenaStanford2016,MSY2016,Gross2017,franz2018mimicking}.


Beyond the rich physics incorporated in the SYK model, the rarity
of solvable, strongly-interacting chaotic systems in quantum
mechanics further highlight its significance. Hence, experimental
realization of the SYK model is worth pursuing. The lack of
experimental quantum simulations of the SYK model nowadays can be
mainly attributed to two facts: it is extremely difficult to
simulate the Hamiltonian with strong randomness and fully nonlocal
fermion interactions, and remains unclear that how to initialize
the simulated system into specific states at different
temperatures and measure the concerned dynamical properties. A
quantum simulator with individual and high-fidelity
controllability will be the key, while the simulation process
should be ``digital"
\cite{georgescu2014quantum,kim2010quantum,luo2016experimental,du2010nmr,kong2016direct,Peng2009}.
As digital quantum simulation often requires error-prone
Trotter-Suzuki decompositions repeatedly, relevant experiments
were still performed on a few qubits
\cite{luo2018experimentally,kandala2017hardware,lv2018quantum,Li:2017fj}.
This is indeed a poignant contrast to current analog quantum
simulation experiments which have already involved about 50
particles \cite{bernien2017probing}, but it should be
understandable that the two approaches are radically different.
Moreover, it is yet impossible to carry out the SYK simulation on
the cloud quantum computing service launched by IBM, as that
service is based on a sequential implementation of elementary
quantum gates rather than dynamical evolution of given
Hamiltonians.

The best route to simulate the SYK model and explore its
fascinating properties at present is via fully controllable
quantum systems, where nuclear magnetic resonance (NMR) is one of
the most suitable systems. The goal of this work is to
experimentally investigate the SYK model, in
particular the fermion pair instability of the SYK NFL and the
chaotic-nonchaotic transition predicted recently~\cite{Xu2017}. We
realized the (0+1)\emph{d} generalized SYK model with $N=8$
Majorana fermions using a four-qubit NMR quantum simulator, and
measured the boson correlation functions at different simulated
temperatures and perturbations. The early-time and late-time decay
behaviors of fermion-pair correlations reflect the fact that there
exist two different phases of the generalized SYK model, i.e.,
maximally chaotic NFL phase and perturbatively weak chaotic
fermion pair condensate phase. The results reveal their
competition under different perturbations, and also the thermal
behavior at different simulated temperatures.

\section*{Results}

\subsection{Generalized SYK model}
The Hamiltonian of $(0+1)d$ generalized SYK model we considered is
given by
\begin{equation}\label{eq:SYK}
H=\frac{J_{ijkl}}{4!}\chi_i\chi_j\chi_k\chi_l+\frac{\mu}{4}C_{ij}C_{kl}\chi_i\chi_j\chi_k\chi_l,
\end{equation}
where $\chi_{i,j,k,l}$ are Majorana fermion operators with
indices $i,j,k,l=1,\cdots,N$, and both $J_{ijkl}$ and $C_{ij}$ are
antisymmetric random tensors drawn from a Gaussian
distribution: $\overline{J_{ijkl}}=0,
\overline{J_{ijkl}^2}=3!J_4^2/N^3$ and $\overline{C_{ij}}=0,
 \overline{C_{ij}C_{kl}}=J^2/N^2(\delta_{ik}\delta_{jl}-\delta_{il}\delta_{jk})$.
Note that $J_4$ has the dimension of energy, while $J$ has the
dimension of (energy)$^{1/2}$. Its phase diagram is shown in Fig.
\ref{fig:phase}a. At $\mu=0$, the Hamiltonian describes the pure
SYK model, whose low temperature state in the limit $N \gg J/T \gg
0$ is a maximally chaotic NFL. As pointed out in
Ref.~\cite{Xu2017}, the SYK fixed point could be unstable towards
fermion pair condensate and spontaneous symmetry breaking, i.e. an
analogue of BCS instability. For instance, a positive $\mu$ term
in the Hamiltonian \eqref{eq:SYK} is a (marginally) relevant
perturbation that drives the spontaneous breaking of the
time-reversal symmetry $\mathcal{T}: \chi_i\to\chi_i, i\to -i$. In
the $\mathcal{T}$-breaking phase, the following bosonic fermion
pair operator
\begin{equation}
b=i C_{ij}\chi_i\chi_j
\end{equation}
develops a persistent correlation $\langle b(t)b(0)\rangle\sim\text{constant}$ that does not decay in time
$t$. We will use this long-time boson correlation as an
experimental signature for the $\mathcal{T}$-breaking phase. The
ordering of $b$ actually has a simple mean-field understanding,
since the $\mu$ term can also be written as $-\mu b^2/2$ which
favors $\langle b \rangle \neq 0$ when $\mu>0$. This can be viewed as
a $(0+1)d$ analog of the Cooper instability of a NFL at low
temperature. In the presence of $\langle b\rangle\neq 0$, the
pairing term $-i\mu\langle b\rangle C_{ij}\chi_i\chi_j$ in the
mean-field Hamiltonian is most relevant at low-energy, which leads
to a non-chaotic ground state in the infrared limit, plus
perturbatively irrelevant interaction that causes weak chaos. On
the other hand, if $\mu$ is negative ($\mu<0$), the spontaneous
symmetry breaking will not be favored and the system will remain
in the maximally chaotic non-Fermi liquid phase.

\subsection{Physical system}


In experiment, we use four spins to simulate $N=8$ Majorana
fermions, as illustrated in Fig. \ref{fig:sample_circuit}a. The Hamiltonian can be encoded into the
spin-1/2 operators via the Jordan Wigner transformation:
\begin{align}\label{eq:JW}
    \chi_{2i-1}=\frac{1}{\sqrt{2}}\sigma_x^1\sigma_x^2\cdots\sigma_x^{i-1}\sigma_z^i,\\ \nonumber
    \chi_{2i}=\frac{1}{\sqrt{2}}\sigma_x^1\sigma_x^2\cdots\sigma_x^{i-1}\sigma_y^i.
\end{align}
Here $\sigma_{x,y,z}$ stand for Pauli matrices. There are 70
$J_{ijkl}$'s, 28 $C_{ij}$'s and four types of spin interactions
(i.e., 1-, 2-, 3- and 4-body interactions) in the case of $N=8$.
The physical system we used has four nuclear spins ($\text{C}_1,
\text{C}_2, \text{C}_3 \text{ and }\text{C}_4$) in the sample of
trans-crotonic acid dissolved in d6-acetone. Its molecular
structure is shown in
Figs. \ref{fig:sample_circuit}b. The natural Hamiltonian of this system in rotating
frame is
\begin{equation}\label{eq:NMR}
\hat{H}_{\text{NMR}}=\sum_{i=1}^{4}\frac{\omega _{i}}{2}\hat{\sigma}
_{z}^{i}+\sum_{i<j,=1}^{4}\frac{\pi J_{ij}}{2}\hat{\sigma} _{z}^{i}\hat{%
\sigma} _{z}^{j},
\end{equation}%
where $\omega _{i}$ represents the chemical shift of spin $i$ and
$J_{ij}$ the coupling constant between spins $i$ and $j$. The relevant Hamiltonian parameters can be seen in Supplementary Information. The
experiment was carried out on a Bruker DRX-700 spectrometer at
room temperature ($T=298$ K). The experiment is divided into three steps: preparation of initial states, simulation of generalized SYK model and measurement of boson correlation functions, as illustrated in Fig.
\ref{fig:sample_circuit}c.

\subsection{Preparation of initial states}
Under high-temperature approximation,
the natural system is originally in the thermal equilibrium state
$\rho_{\text{eq}}\approx(\mathbb{I}+\epsilon\sum_{i=1}^4\sigma_i^z)/2^4$,
where $\mathbb{I}$ is the identity and $\epsilon\sim10^{-5}$ is
the polarization. During our quantum computation, the evolution
preserves the unit operator $\mathbb{I}$, so we omit it and
rewrite $\rho_{eq}=\epsilon\sum_{i=1}^4 \sigma_i^z$. Hereinafter
we used the deviation density matrices as
`states'~\cite{Chuang1998}. Starting from $\rho_{\text{eq}}$, the
system was prepared into the initial `states':
$\rho_i^{\text{Real}}=(\rho_{\text{eq}}^{H}b+b\rho_{\text{eq}}^{H})/2$
and
$\rho_i^{\text{Imag}}=-i(\rho_{\text{eq}}^{H}b-b\rho_{\text{eq}}^{H})/2$,
where $\rho_{\text{eq}}^H=e^{-\beta H}/\text{Tr}(e^{-\beta H})$.
These initial states can be implemented, as shown in Fig. \ref{fig:sample_circuit}c. The network with single-qubit rotations and free evolutions of
the natural Hamiltonian allow us to get the states (before
the first $z$-direction gradient field), whose diagonal elements
equal to the eigenvalues of $\rho_i$. The rotation angles
$\theta_j$'s for different $\beta$ and $H$ were given in the Supplementary
Information. The CNOT gates were applied to remove zero quantum
coherence that cannot be averaged out by the $z$-direction
gradient fields. The states after the third $z$-direction gradient
field are thus the diagonal density matrices, i.e.,
$\rho_i^d=V^{\dag}\rho_iV$, where $V$ is the basis transformation
between computational basis and eigenvectors of $\rho_i$. When
performing the $V$ transformation, we can obtain the initial
states $\rho_i$.

\subsection{Simulation of generalized SYK model}
The evolution of generalized SYK model can be simulated with a controllable NMR system effciently, as pointed out originally by Feynman
\cite{Feynman1982,Solano2017}. We rewrite the Hamiltonian (\ref{eq:SYK}) as the sum of spin interactions, i.e.,
\begin{equation}
	H=\sum_{s=1}^{70} H_s=\sum_{s=1}^{70}a_{ijkl}^s\sigma_{\alpha_i}^1\sigma_{\alpha_j}^2\sigma_{\alpha_k}^3\sigma_{\alpha_l}^4,
\end{equation}
according to equation \eqref{eq:JW}, where subscripts 
$\alpha=\{0,x,y,z\}$ label the corresponding Pauli matrices, and $\sigma_0=\mathbb{I}$. All random coefficients $a_{ijkl}^s$
are shown in Supplementary Information. Using the Trotter-Suzuki formula, its exact time evolution operator can
be decomposed into \cite{Lloyd1996},
\begin{equation}\label{eq:DQS}
e^{-i H\tau}=\left(\prod_{s=1}^{70} e^{-i H_s \tau/n}\right)^n+\sum_{s<s'}\frac{[H_s,H_{s'}]\tau^2}{2n}+O(|a|^3\tau^3/n^2),
\end{equation}
where $|a|=\left(\overline{|a_{ijkl}^s|^2}\right)^{-1/2}\approx 0.64$
for $\mu=\pm 5$, and $\approx 0.27 $ for $\mu=0$ (Here $J=\sqrt{J_4}=1$ were chosen in experiments). Obtaining this exact time evolution is a difficult problem to deal with a quantum simulation, but it is possible to handle the first-order product operator $\left(\prod_{s=1}^{70} e^{-i H_s \tau/n}\right)^n$. The approximate simulation of $e^{-i\mathcal{H}\tau}\approx \left(\prod_{s=1}^{70} e^{-i H_s \tau/n}\right)^n$  can take place to within a desired accuracy by choosing sufficiently large $n$. In particular, if
$[H_s,H_{s'}]=0$, there is a boost in accuracy. The fidelity between $e^{-i H\tau}$ and $\left(\prod_{s=1}^{70} e^{-i H_s \tau/n}\right)^n$ as a
function of $\tau$ and $n$ is shown in Fig.
\ref{fig:fidelity_interaction}a. For example, when
$\text{ln}(\tau)=2$ and $\text{log}(n)=1.55$, the fidelity is over
$0.99$. 

For simulating the $\left(\prod_{s=1}^{70} e^{-i H_s \tau/n}\right)^n$, we evolve the system forward locally over small, discrete time slices, i.e.,  $e^{-iH_1\tau/n}, e^{-iH_2\tau/n}$, and so on, up to $e^{-iH_{70}\tau/n}$, and repeat $n$ times. Each local many-body spin interaction of $H_s=a_{ijkl}^s\sigma_{\alpha_i}^1\sigma_{\alpha_j}^2\sigma_{\alpha_k}^3\sigma_{\alpha_l}^4$ can be effectively created by the means of coherent control acting on the physical system of nuclear spins \cite{Cory1999,Laflamme2005,Peng2009,Luo2016,Li:2017fj,Liu2008}.  The task in coherent control is to design a pulse sequence for finding the appropriate amplitudes and phases of radio-frequency (RF) fields. To improve the control performance in our experiment, we employed the gradient ascent pulse engineering (GRAPE) algorithm \cite{Glaser2005} to optimize the field parameters of a shaped pulse. The shaped pulse with the duration of 100 ms and the slices of 4000 was designed to have theoretical fidelity over 0.99, and be robust against the $5\%$ inhomogeneity of RF fields. The detail of experimental simulation and the shaped pulse can be seen in Supplementary Information.

It is necessary to note that the quantum simulation algorithm is efficient. 
 As shown in Fig. \ref{fig:fidelity_interaction}b, a $k$-body spin interaction with the form of $\sigma_z^1\sigma_z^2\cdots\sigma_z^k$ can be decomposed as a $(k-1)$-body interaction by the following iteration,
\begin{equation}\label{eq:MBS}
    e^{-i\frac{\pi}{2}\sigma_z^1\sigma_z^2\cdots\sigma_z^k \tau}=P_1e^{-i\sigma_z^2\cdots\sigma_z^k\tau}P_2,
\end{equation}
where
$P_1=e^{-i\pi\sigma_x^{2}/4}e^{-i\pi\sigma_z^{1}\sigma_z^{2}/4}e^{-i\pi\sigma_y^{2}/4}$,
and $
P_2=e^{-i\pi\sigma_y^{2}/4}e^{-i\pi\sigma_z^{1}\sigma_z^{2}/4}e^{i\pi/2\sigma_y^{2}}e^{i\pi\sigma_x^{2}/4}$.
For a $k$-body $(k>2)$ interaction, it requires $5(k-2)$ 1-body
interactions and $2(k-2)$ 2-body interactions. Given an accuracy $\epsilon$, the total number
of gates is $n\sum_{i=1}^ml(k)$, where $n\propto
|a|^2\tau^2/\epsilon$, $m=\binom{N}{4}$ is the number of spin
interactions, and $l(k)=7(k-2)$ counts the number of
gates in implementing a $k$-body $(k\leq N/2)$ spin interaction.
Therefore, the total number of gates $\propto
|a|^2\tau^2N^5/\epsilon$ grows polynomially with the number of
Majorana fermions $N$, indicating that digital quantum simulation
of the generalized SYK model is efficient. 


\subsection{Measurement of boson correlation function}

Finally, we measure the boson correlation function to probe the
instability of the SYK non-fermi-liquid ground state towards
spontaneous symmetry breaking at different values of $\beta$ and
$\mu$. The boson correlation function is defined as
\begin{equation}\label{eq:CF}
\langle b(\tau)b(0)\rangle_{\beta}=\frac{\text{Tr}(e^{-\beta H}
e^{-i H\tau}b e^{i H\tau} b)}{\text{Tr}(e^{-\beta H})}.
\end{equation}
To remove its initial value fluctuation from sample to sample
(here a sample means that we randomly generate a different
Hamiltonian), we average the normalized correlation function over
random samples,
\begin{equation}\label{eq:NCF}
\overline{|D(\tau)|}=\text{avg}\left(\left|
\frac{\langle b(\tau)b(0)\rangle_{\beta}}{\langle b(0)b(0)\rangle_{\beta}}\right|\right),
\end{equation}
where the normalization is applied to avoid the unphysical phase
interference among different samples. In experiment, we randomly generated eight samples or Hamiltonians, as shown in supplementary information. 

Starting from initial states
$\rho_i^{\text{Real}}$ and $\rho_i^{\text{Imag}}$, the real and
imaginary parts of $\langle b(\tau)b(0)\rangle_{\beta}$ can be
obtained by measuring the bosonic fermion pair operator $b$, namely,
$\text{Re}(\langle b(\tau)b(0)\rangle_{\beta})=\text{Tr}(e^{-iH\tau}\rho_i^{\text{Real}}e^{iH\tau}b)$,
and $\text{Im}(\langle b(\tau)b(0)\rangle_{\beta})=\text{Tr}(e^{-iH\tau}\rho_i^{\text{Imag}}e^{iH\tau}b)$.
In NMR, the measured signal via quadrature detection is given by
\cite{LEE2002349},
\begin{equation}\label{eq:NMRsignal}
    S(t)=\text{Tr}\left[\rho_f M^{\dag}e^{i{H}_{\text{NMR}}t}\sum_{j=1}^4(\sigma_x^j+i\sigma_y^j)e^{-i{H}_{\text{NMR}}t}M\right],
\end{equation}
where $\rho_f=e^{-iH\tau}\rho_i^{\text{Real}}e^{iH\tau}$ or $e^{-iH\tau}\rho_i^{\text{Imag}}e^{iH\tau}$ is the output density matrix, and $M$ represents a series of readout operators. We can see that the NMR signal consists of both
real and imaginary parts, and is the average of transverse magnetization without any readout pulse. The bosonic fermion pair
operator $b$ including 28 spin operators can be obtained by
designing a specific set of readout pulses. To get all spin operators of $b$, i.e., $\text{Tr}[\rho_f
b]$ completely, we used five readout pulses in experiments. The readout pulses and their corresponding readout spin operators
are listed in Table 1.


\subsection{Experimental results}

The main experimental results are
shown in Fig. \ref{fig:result}, which were obtained by averaging over eight random samples. 
The data for each random sample is available in Supplementary Information.
The error bars mainly come from
the fitting procedure (about $1\%$) and fluctuation of random
samples, which is  less than $2\%$ when $\text{ln}(\tau)\leq 1$ and
around $15\%$ when $\text{ln}(\tau)> 1$. After normalization of the
correlation function to compensate for the effect of decoherence,
the experimental result is in good agreement with theoretical
predictions. 

Let us first look at the low temperature  result ($\beta=20$) in
Fig. \ref{fig:result}c. The boson correlation was measured for
three different values of $\mu$. For both $\mu=0$ and $\mu=-5$,
the boson correlations decay quickly following the similar
manner, which means that they are relevant chaotic phases. Because the low temperature state at $\mu=0$ described by the pure SYK model is a maximally chaotic NFL phase. While for $\mu=5$, the boson correlation decays much
slower and saturates to a relatively large value. This difference
indicates the long-time order of $\mu>0$
is the spontaneous $\mathcal{T}$-breaking phase. In contrast, for
$\mu<0$, there is no such instability towards symmetry breaking.
So by changing the sign of $\mu$, the system goes through a continuous chaotic-nonchaotic transition, whose critical properties are analogous to that of the Kosterlitz-Thouless transition \cite{Xu2017}.

The fact that boson correlation still saturates to some finite
value in the NFL phase for $\mu\leq0$ is due to the
finite size of our system. Theoretically, in the NFL
phase ($\mu\leq0$), the saturate value of boson correlation decays
towards zero with the growth of the system size. In the symmetry
breaking phase ($\mu>0$), the saturate value scales towards a
finite value in the thermodynamic limit. Numerical simulations of
this scaling behavior is provided in Supplementary Information. In
spite of the finite size, different phases of the generalized SYK
model are indeed demonstrated by different behaviors of the boson
correlation in our experiment.

\begin{figure}
\begin{center}
\includegraphics[width=1\linewidth]{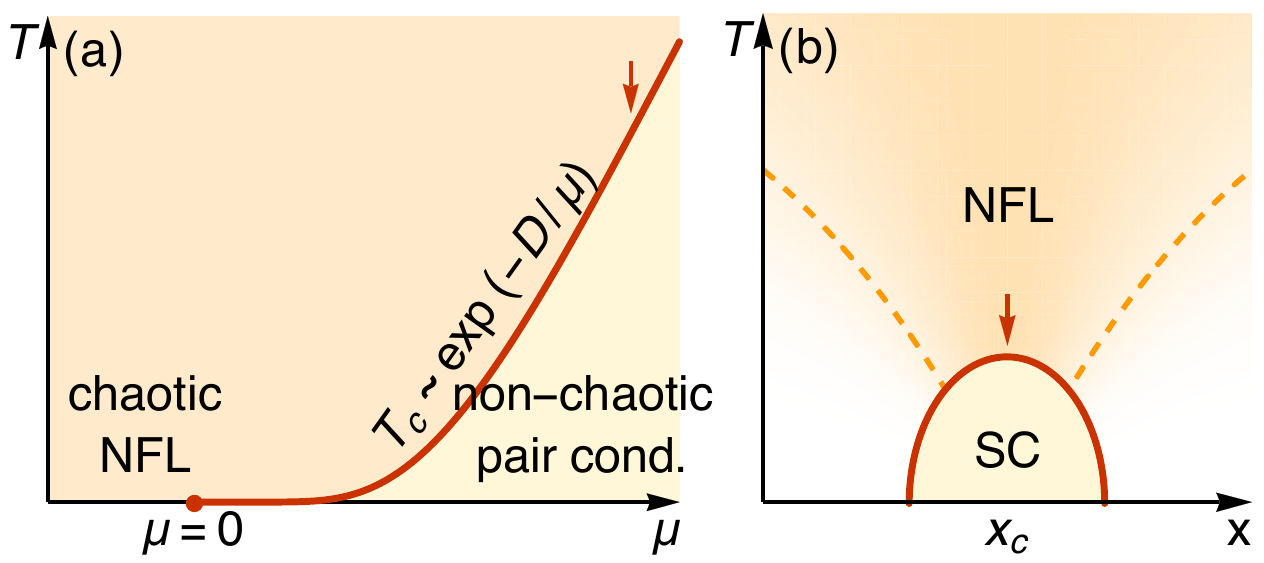}\\
\caption{{\bf{Schematic phase diagram of a,}} the generalized
SYK model in Eq.\,\eqref{eq:SYK} with pair condensation
instability on the $\mu>0$ side, and {\bf{b,}} the standard
non-Fermi liquid (NFL) behavior in the proximity of a quantum
critical point covered by a superconducting (SC) dome at low
temperature.} \label{fig:phase}
\end{center}
\end{figure}

\begin{figure}
\begin{center}
\includegraphics[width=0.95\linewidth]{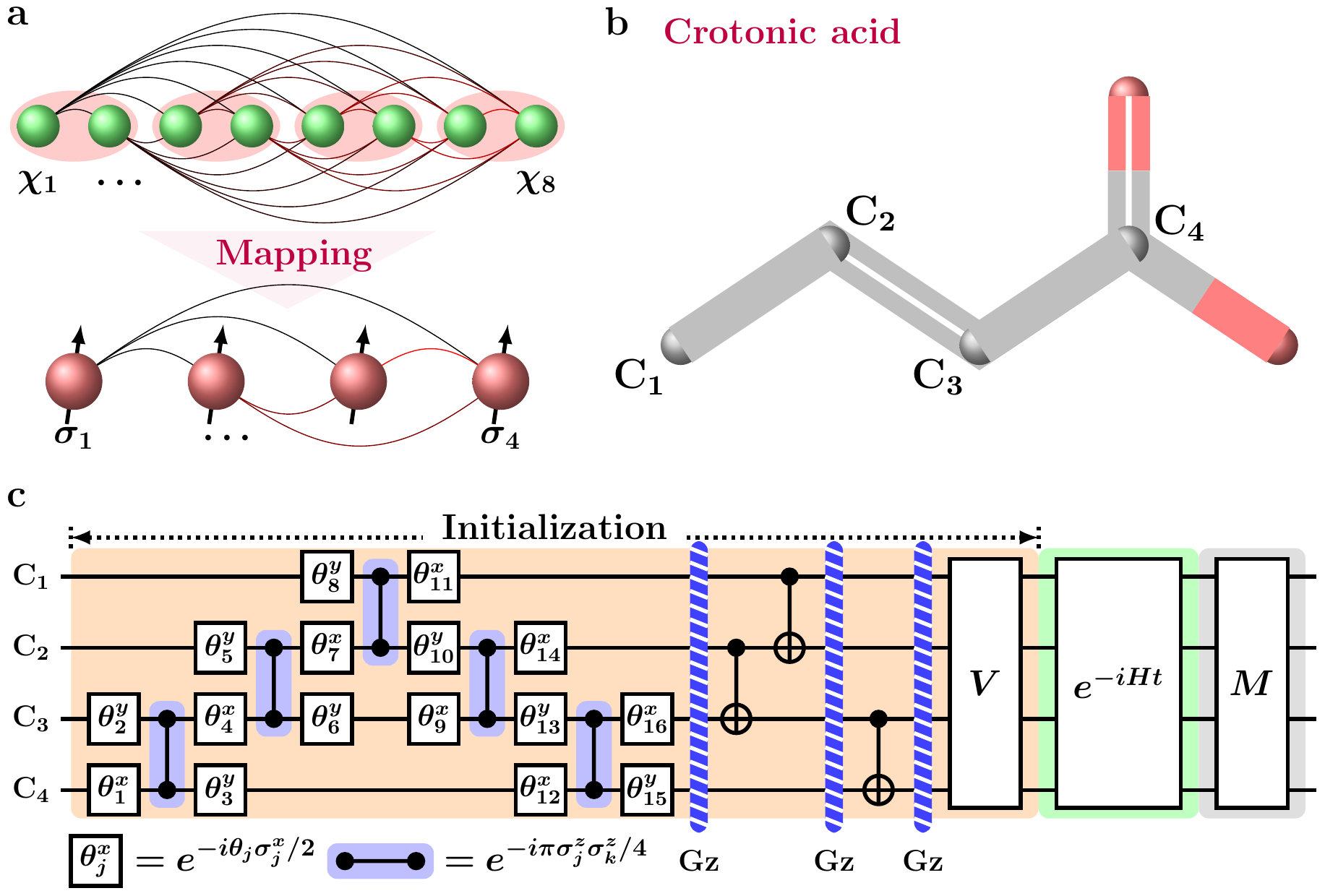}\\
\caption{{\bf{Scheme for experimentally simulating generalized SYK model and measuring the boson correlation function. a,}} The generalized SYK model with 8 Majorana fermions is mapped onto a four spin model. The curved lines denote the Majorana fermion-fermion or spin-spin interactions. {\bf{b,}} Molecular structure of $^{13}$C-labeled trans-crotonic acid, where nuclear spins of
$\text{C}_1, \text{C}_2, \text{C}_3 \text{ and }\text{C}_4$
are used as a four-qubit quantum simulator. All
protons are decoupled throughout the experiments. 
{\bf{c,}} Quantum circuit for
measuring the boson correlation function. $V$ is the basis
transformation from the computational basis to the eigenvectors of
initial states $\rho_i$. $M$ represents five readout pulses for
observing $b$ operator.}\label{fig:sample_circuit}
\end{center}
\end{figure}

\begin{figure}
\begin{center}
\includegraphics[width=0.9\linewidth]{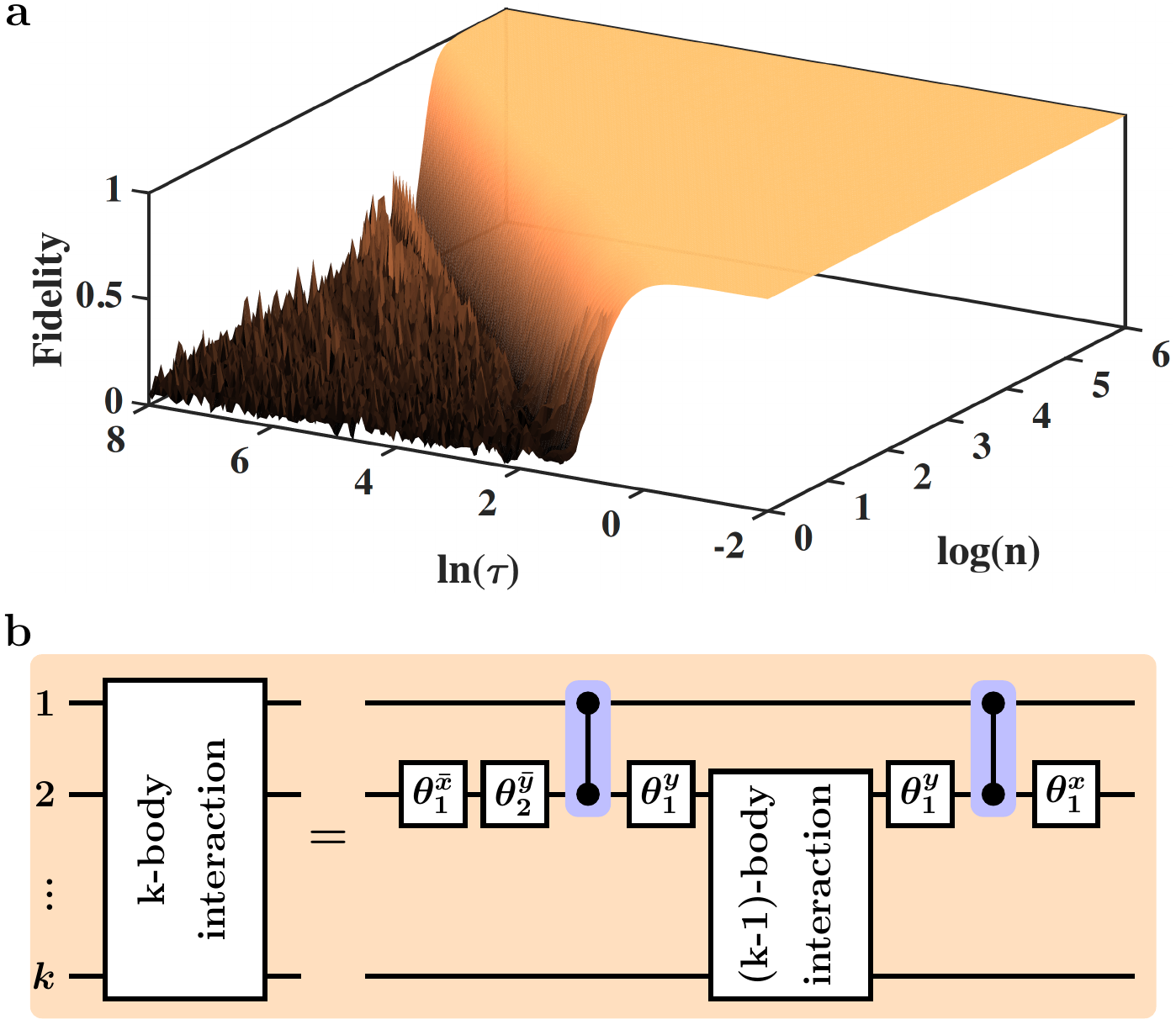}
\caption{{\bf{Fidelity of the Trotter-Suzuki decomposition and pulse sequence for implementing a $k$-body interaction.}} In {\bf{a}}, the fidelities are calculated between $e^{-i\mathcal{H}\tau}$ and its decomposition $\left(\prod_{s=1}^{70} e^{-i \mathcal{H}_s \tau/n}\right)^n$ as a function of $\tau$ and $n$. In {\bf{b}}, to simulate a $k$-body interaction, the pulse sequence includes $5(k-2)$ 1-body interactions and $2(k-2)$ 2-body interactions. $\theta_1=\pi/2$ and $\theta_2=\pi$.}\label{fig:fidelity_interaction}
\end{center}
\end{figure}

\begin{figure}
\begin{center}
\includegraphics[width=0.5\linewidth]{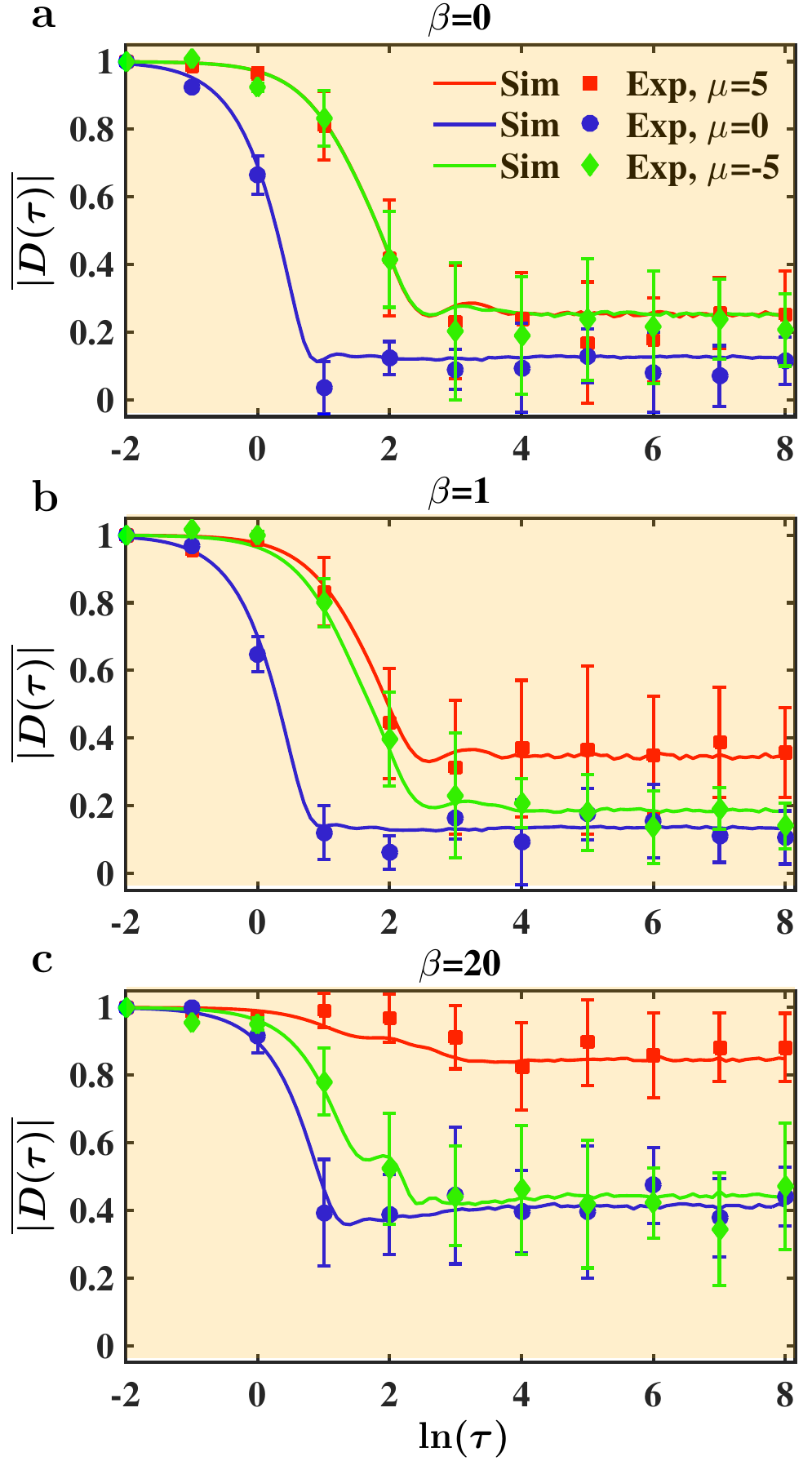}
\caption{{\bf{Boson correlation functions for different
$\beta$ and $\mu$.}} Solid lines are simulation results. Points are experimental data obtained by averaging over eight random samples. In {\bf{a}}, there is no significant difference of boson correlation functions between the $\mu=\pm 5$ cases. But this difference grows as the temperature ($T=1/\beta$) goes down, as shown in {\bf{b}}. At low temperature of {\bf{c}},  the decay behavior of boson correlations for $\mu=-5$ becomes similar with the case of $\mu=0$. For $\mu=0$, the pure SYK model at low temperature describes a maximally chaotic NFL phase with the fastest decay rate of boson correlation. While for $\mu=5$, the boson correlation decays much slower and saturates to a relatively large value, which corresponds to a spontaneous $\mathcal{T}$-breaking phase.}\label{fig:result}
\end{center}
\end{figure}

As we raise the temperature to $\beta=1$ in Fig.
\ref{fig:result}b, the boson condensation is destroyed by the
thermal fluctuation and the long-time correlation is suppressed.
In the thermodynamic limit, the transition temperature scales as
$T_c\sim\text{exp}(-\sqrt{\pi}J_4/2J^2\mu)$ \cite{Xu2017}
illustrated in Fig. \ref{fig:phase}a. At infinite temperature
($\beta=0$) in Fig. \ref{fig:result}a, the $\mu=\pm5$ curves
coincide, since the boson correlation $D(\tau)=\text{Tr}(e^{i H \tau} b
e^{-i H \tau}b)$ in this scenario is invariant under the
transformation $H\to-H$. We observe the slightly different
behaviors of boson correlations between the $\mu\neq0$ and $\mu=0$
cases. The correlation decays fastest and to the lowest saturation
value at $\mu=0$, which is consistent with the fact that the pure
SYK model is maximally chaotic scrambling the order parameter most
thoroughly.


\section*{Discussion}
In summary, we report the experimental realization of the
SYK model and its generalization. The measurements of fermion-pair
correlation functions in our experiment exhibit the instability of the maximally
chaotic NFL phase of the SYK model against certain types of
four-fermion perturbations, which drives the system into a less
chaotic fermion pair condensed phase with spontaneous
$\mathcal{T}$-breaking.
These successful experimental
demonstrations rely heavily on the fully controllability of our NMR
quantum simulator. The NMR system has the advantages of well characterized qubits, long decoherence time, and fine control of nuclear spins through RF fields, which enable us to simulate the dynamics of generalized SYK model.
Our experiment demonstrates the first step towards quantum
simulation of non-Fermi-liquid states in strongly interacting systems. The methods used here
can also be adapted in other quantum platforms, and may provide a
new path towards exploring the holographic duality. It will also be interesting to further test non-equilibrium dynamic and the scrambling of information by measuring the out-of-time-order correlation, which has been proposed as a identification of chaos in quantum systems. 

One major concern is about the scalability of the control
techniques adopted in the experiment, as it is supposed to be the
largest obstacle when performing higher-dimensional digital
quantum simulations. In fact, the gradient-based optimal control
is the bottleneck that limits future experimental size. Despite
its extraordinary performance in small number of qubits, this
technique does not posses well scalability in principle. Recently,
an alternative method that utilizes the power of quantum processor
together with machine learning techniques to enhance quantum
control was reported \cite{li2017hybrid,lu2017enhancing}. This
method is also efficient, i.e., requires polynomial time for
optimization with the number of qubits. Improvement of control
fidelities was solidly demonstrated on a 12-qubit system
\cite{lu2017enhancing}. Compared to the results here, this
technique leads to similar control accuracies according to our
numerical simulation. As this new optimization method can be
scaled up to many qubits, we anticipate it to underpin future
quantum simulation tasks of more complex SYK as well as other
models. For instance, one of most exciting prospects is to mimic the black holes and thus experimentally test the quantum gravitation ideas in the laboratory. 

\section*{Data Availability} The datasets generated during and/or analysed during the current study are available from the corresponding author on reasonable request.

\begin{addendum}

 \item We thank X. Peng, L. Hung and J. Mei for helpful discussion. This work was supported by CIFAR, NSERC and Industry of Canada. Z.L. acknowledge the support from the National Natural Science Foundation of China (Grants No. 11805008, No. 11734002 and No. 11374032). C.X. is supported by the David and Lucile Packard Foundation and NSF Grant No. DMR-1151208. C.J. is partly supported by the Gordon and Betty Moore Foundations EPiQS Initiative through Grant GBMF4304. D. L. are supported by the National Natural Science Foundation of China
(Grants No. 11605005, No. 11875159 and No. U1801661),Science, Technology and Innovation
Commission of Shenzhen Municipality (Grants No. ZDSYS20170303165926217 and No.
JCYJ20170412152620376), Guangdong Innovative and Entrepreneurial Research Team Program
(Grant No. 2016ZT06D348).

 \item[Author Contributions] Z.L. designed and performed the experiment. Y.Y., C.X., and C.J. developed the theory. D.L., J.L., B.Z., and R.L. supervised the project. Z.L., Y.Y., D.L. and C.X. wrote the draft. All authors contributed to discussing the results and writing the manuscript.
 \item[Competing Interests] The authors declare that they have no competing interests.
 \item[Correspondence] Correspondence and requests for materials
should be addressed to D. W. L. (ludw@sustech.edu.cn) or B. Z. (zengb@uoguelph.ca).
\end{addendum}

\clearpage

\clearpage
\begin{table*}
\caption{Readout pulses and their corresponding readout spin operators. $R_{x, y}^j$ is the notation of the $j_{\text{th}}$-spin rotation by $\pi/2$ about
$x$ or $y$ axis. X, Y, and Z are Pauli matrices, while I is the unit operator.}
\label{tab:readout}
  \centering
  \begin{tabular}{cc}
  \hline\hline
   Readout pulses & Readout operators \\
   \hline
   $R_x^3$ & XIII, YZII, ZYII, IXII, IYXZ, IZXY, IZYI, IZZI, IIYZ, IIZZ, IIIX\\
   $R_x^2R_y^3$ & YYII, ZXXZ, ZZII, IYXY, IYYI, IYZI, IZXZ\\
   $R_y^2R_y^3$ & YXXZ, ZXXY, ZXYI, ZXZI\\
   $R_x^1R_y^2R_x^4$ & YXXY, YXYI, IIXI, IIYY\\
   $R_x^1R_y^2R_x^3R_x^4$ & YXZI, IIZY\\
   \hline\hline
  \end{tabular}
\end{table*}

\clearpage
\setcounter{figure}{0}
\setcounter{equation}{0}
\setcounter{table}{0}
\renewcommand{\thefigure}{S\arabic{figure}}
\renewcommand{\theequation}{S\arabic{equation}}
\renewcommand{\thetable}{S\arabic{table}}

\begin{center}
	\textbf{Supplementary Information for ``Quantum Simulation of the non-Fermi-Liquid State of Sachdev-Ye-Kitaev Model''}
\end{center}

\begin{figure}[!htp]
\begin{center}
\includegraphics[width=0.95\linewidth]{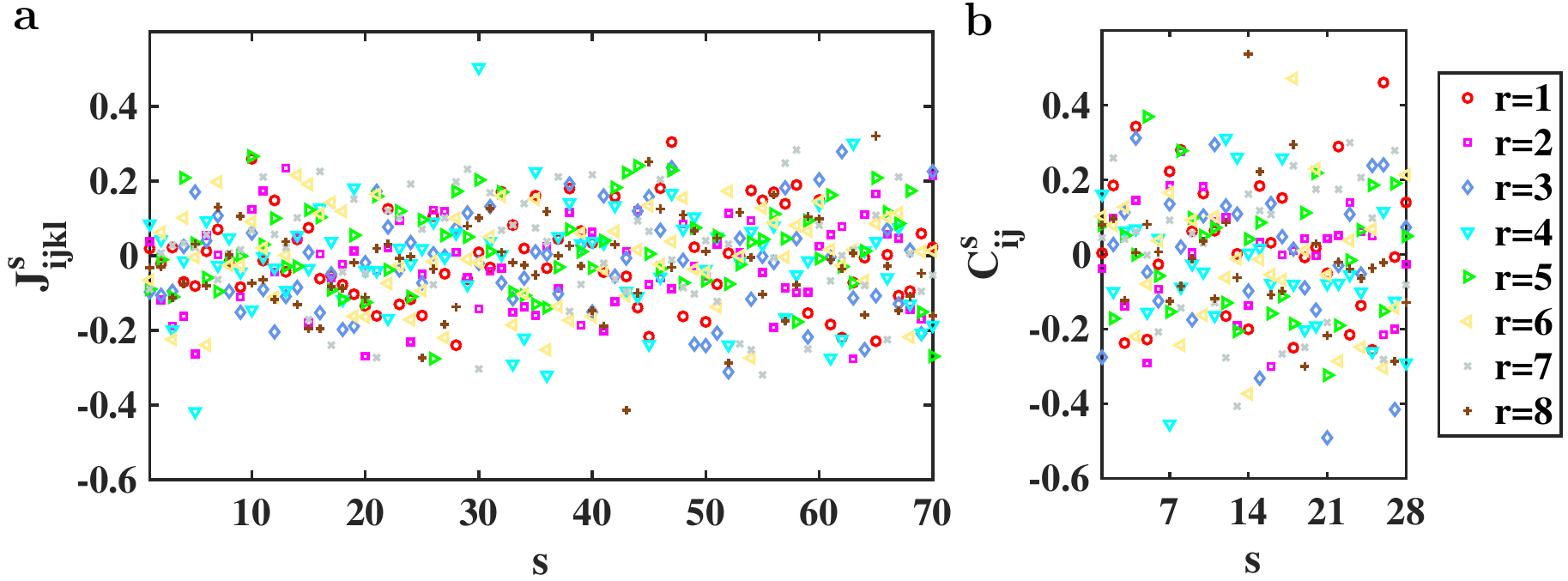}
\caption{Antisymmetric random Tensors of $J_{ijkl}$ and $C_{ij}$ in different random Hamiltonians.}
\label{fig:JC}
\end{center}
\end{figure}

\textbf{1. Hamiltonian parameters of the generalized SYK model.}
The Hamiltonian of $(0+1)d$ generalized SYK model with $N=8$ Majorana fermions is given by
\begin{equation}\label{eq:SYK}
H=\frac{J_{ijkl}}{4!}\chi_i\chi_j\chi_k\chi_l+\frac{\mu}{4}C_{ij}C_{kl}\chi_i\chi_j\chi_k\chi_l.
\end{equation}
The antisymmetric random tensors of $J_{ijkl}$ and $C_{ij}$ are drawn from the Gaussian distribution: $\overline{J_{ijkl}}=0,
\overline{J_{ijkl}^2}=3!J_4^2/N^3$ and $\overline{C_{ij}}=0,
 \overline{C_{ij}C_{kl}}=2J^2/N^2(\delta_{ik}\delta_{jl}-\delta_{il}\delta_{jk})$ (here $J=\sqrt{J_4}=1$),  and plotted in Figs. \ref{fig:JC}(a) and \ref{fig:JC}(b), respectively. For $N=8$, there are 70 $J_{ijkl}$s and 28 $C_{ij}$s in each sample or a random Hamiltonian. In experiments, we randomly generated $r=1,2,\dots,8$ different Hamiltonians. 
 
\begin{figure}
\begin{center}
\includegraphics[width=0.95\linewidth]{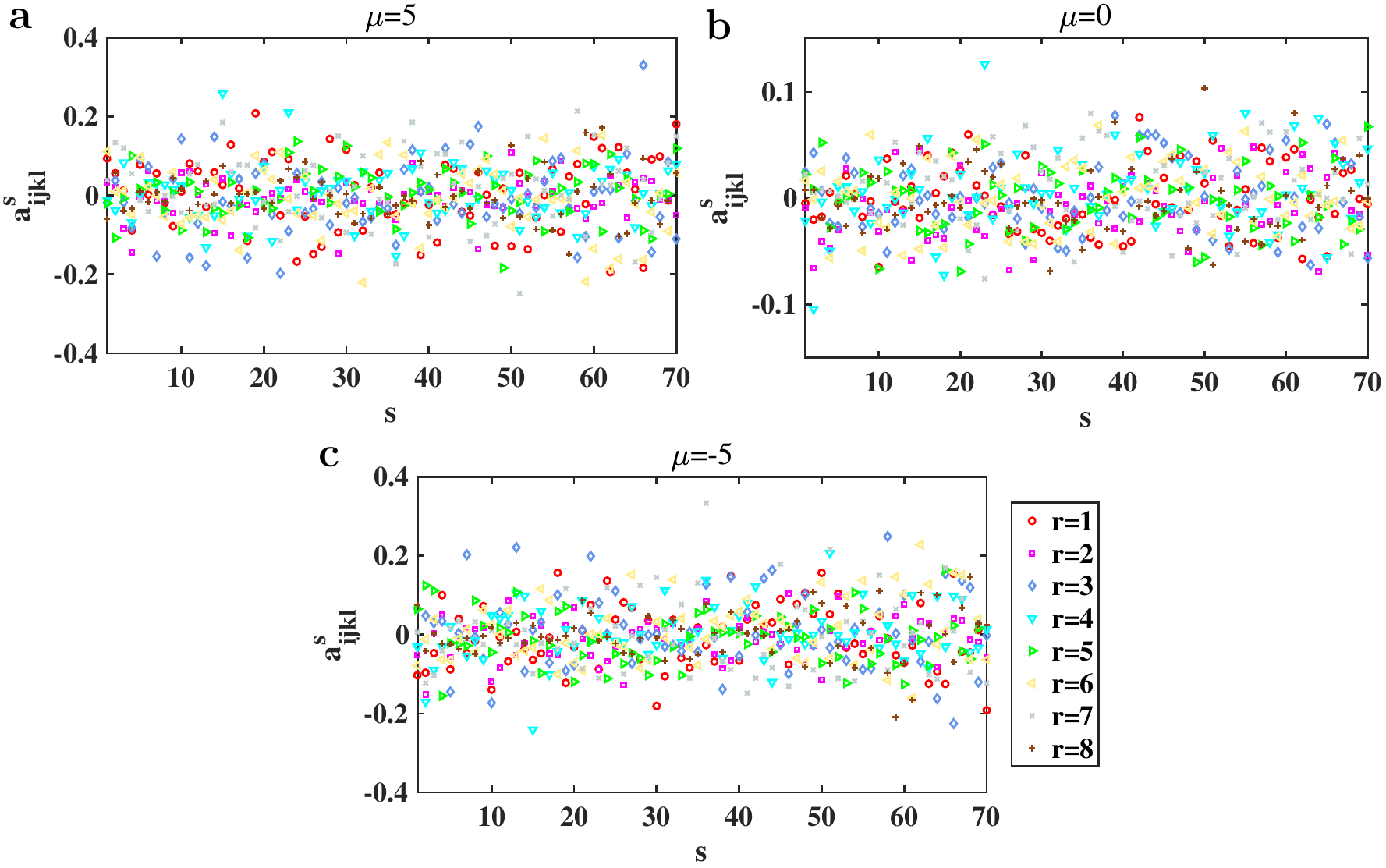}
\caption{Random coefficients of $a_{ijkl}^s$ for different $\mu$.}
\label{fig:aijkl}
\end{center}
\end{figure}

Using the Jordan-Wigner transformation, Hamiltonian (\ref{eq:SYK}) can be rewritten as the sum of spin interactions,
 \begin{equation}\label{eq:Hs}
 	H=\sum_{s=1}^{70}H_s=\sum_{s=1}^{70}a_{ijkl}^s\sigma_{\alpha_i}^1\sigma_{\alpha_j}^2\sigma_{\alpha_k}^3\sigma_{\alpha_l}^4,
 \end{equation}
where subscripts $\alpha=\{0,x,y,z\}$ label the corresponding Pauli matrices, and $\sigma_0=\mathbb{I}$. All subscripts of spin interactions are listed in Table. \ref{tab:spin interaction}. For example, the first term of $xx00$ in Table. \ref{tab:spin interaction} represents the 2-body spin interaction, i.e., $\sigma_x^1\sigma_x^2$. The random coefficients of $a_{ijkl}^s$ for different $\mu$ are shown in Fig. \ref{fig:aijkl}.

\begin{table*}[!htp]
\caption{The subscripts for denoting the spin interactions in Hamiltonian (\ref{eq:Hs}).}
\label{tab:spin interaction}
  \centering
  \begin{tabular}{cccccccccccccc}
  \hline\hline
   $xx00$ & $xyxy$ & $xyxz$ & $xyy0$ & $xyz0$ & $xzxy$  & $xzxz$ & $xzy0$ & $xzz0$ & $x0x0$ & $x0yy$ & $x0yz$ & $x0zy$ & $x0zz$\\   
 
    $x00x$ & $yxyx$ & $yxzx$ & $yx0y$ & $yx0z$ & $yyx0$  & $yyyy$ & $yyyz$ & $yyzy$ & $yyzz$ & $yy0x$ & $yzx0$ & $yzyy$ & $yzyz$\\
   
    $yzzy$ & $yzzz$ & $yz0x$ & $y0xy$ & $y0xz$ & $y0y0$  & $y0z0$ & $zxyx$ & $zxzx$ & $zx0y$ & $zx0z$ & $zyx0$ & $zyyy$ & $zyyz$\\
  
    $zyzy$ & $zyzz$ & $zy0x$ & $zzx0$ & $zzyy$ & $zzyz$  & $zzzy$ & $zzzz$ & $zz0x$ & $z0xy$ & $z0xz$ & $z0y0$ & $z0z0$ & $0xx0$\\
         
    $0xyy$ & $0xyz$ & $0xzy$ & $0xzz$ & $0x0x$ & $0yyx$  & $0yzx$ & $0y0y$ & $0y0z$ & $0zyx$ & $0zzx$ & $0z0y$ & $0z0z$ & $00xx$\\
    \hline\hline   
  \end{tabular}
\end{table*}

\textbf{2. Relevant parameters of the nuclear spin system.} 
The physical system used in the experiment consists of four carbon-13 nuclear spins of  trans-crotonic acid. Its relevant parameters including chemical shifts, J-couplings and relaxation times are shown in Table. \ref{tab:sample parameter}.

\begin{table}
\caption{The Hamiltonianl parameters of trans-crotonic acid. Diagonal and off-diagonal elements
represent the chemical shifts and $J$-coupling constants (in Hz),
respectively. The measured spin-lattice relaxation times $T_1$ (in
seconds) and spin-spin relaxation times $T_2$ (in seconds) are
shown in the last two columns.}
\label{tab:sample parameter}
  \centering
  \begin{tabular}{c|cccccc}
         & $\mathbf{C_1}$ & $\mathbf{C_2}$ & $\mathbf{C_3}$ & $\mathbf{C_4}$ & $\bm{T_1(s)}$ & $\bm{T_2(s)}$ \\
    \hline
    $\mathbf{C_1}$ & 2989 &       &       &       & 5.7   &  1.02\\
    $\mathbf{C_2}$ & 41.6 & 25459 &       &       & 5.3   &  0.92\\
    $\mathbf{C_3}$ & 1.4  & 69.7  & 21592 &       & 5.6   &  0.89\\
    $\mathbf{C_4}$ & 7.0  & 1.2   & 72.2  & 29341 & 10.2  &  0.94
  \end{tabular}
\end{table}

\begin{figure*}[!htp]
	\includegraphics[width=1\linewidth]{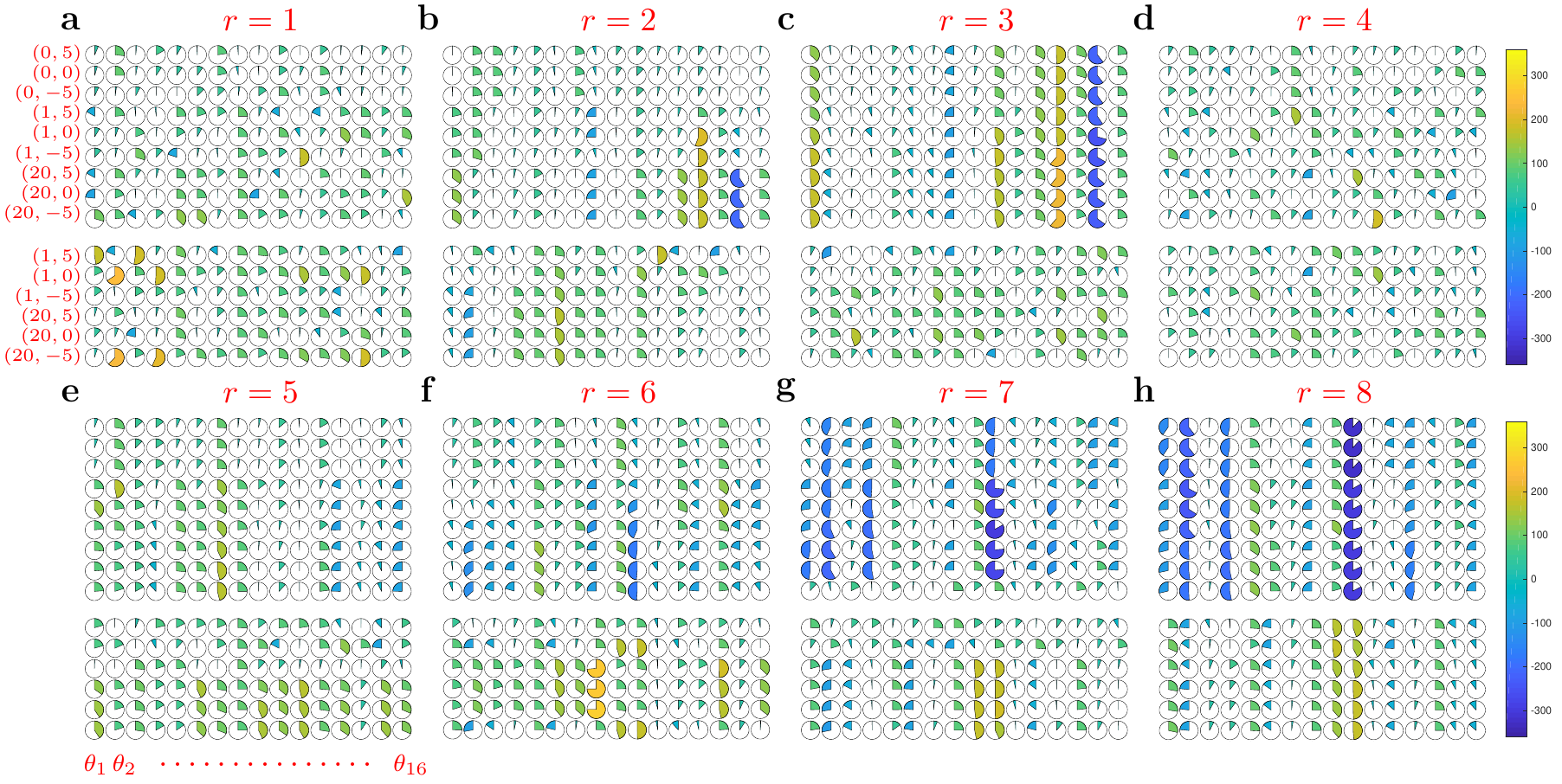}
	\caption{The single-qubit rotation angles (degree unit) for  preparing initial states $\rho_i(\beta,\mu)$ in different random Hamiltonians, where $\beta=0, 1, 20$ and $\mu=5, 0, -5$. The top and bottom panels are used to prepare $\rho_i^{\text{Real}}(\beta,\mu)$ and $\rho_i^{\text{Imag}}(\beta,\mu)$, respectively. Because $\rho_i^{\text{Imag}}(0,\mu)=0$, there is no rotation angle listed here. }
	\label{fig:rotationangles}
\end{figure*}

\textbf{3. Rotation angles for preparing initial states.} The initial 'states' that need to be prepared in our experiments are $\rho_i^{\text{Real}}=(\rho_{\text{eq}}^{H}b+b\rho_{\text{eq}}^{H})/2$
and
$\rho_i^{\text{Imag}}=-i(\rho_{\text{eq}}^{H}b-b\rho_{\text{eq}}^{H})/2$,
where $\rho_{\text{eq}}^H=e^{-\beta H}/\text{Tr}(e^{-\beta H})$. Given a random Hamiltonian and  temperature $\beta$, the sixteen single-qubit rotations and free evolutions of nature NMR Hamiltonian enable the system to be prepared into the specific states, i.e., their diagonal elements of density matrices equal to the eigenvalues of these initial 'states' . The single-qubit rotation angles for preparing $\rho_i^{\text{Real}}(\beta,\mu)$ and $\rho_i^{\text{Imag}}(\beta,\mu)$ in different random Hamiltonians are shown in Fig. \ref{fig:rotationangles}. When $\beta=0$, $\rho_i^{\text{Imag}}(0,\mu)=0$ for any $\mu$, and is not necessary to be prepared.

\begin{figure}[!htp]
\begin{center}
\includegraphics[width=0.8\linewidth]{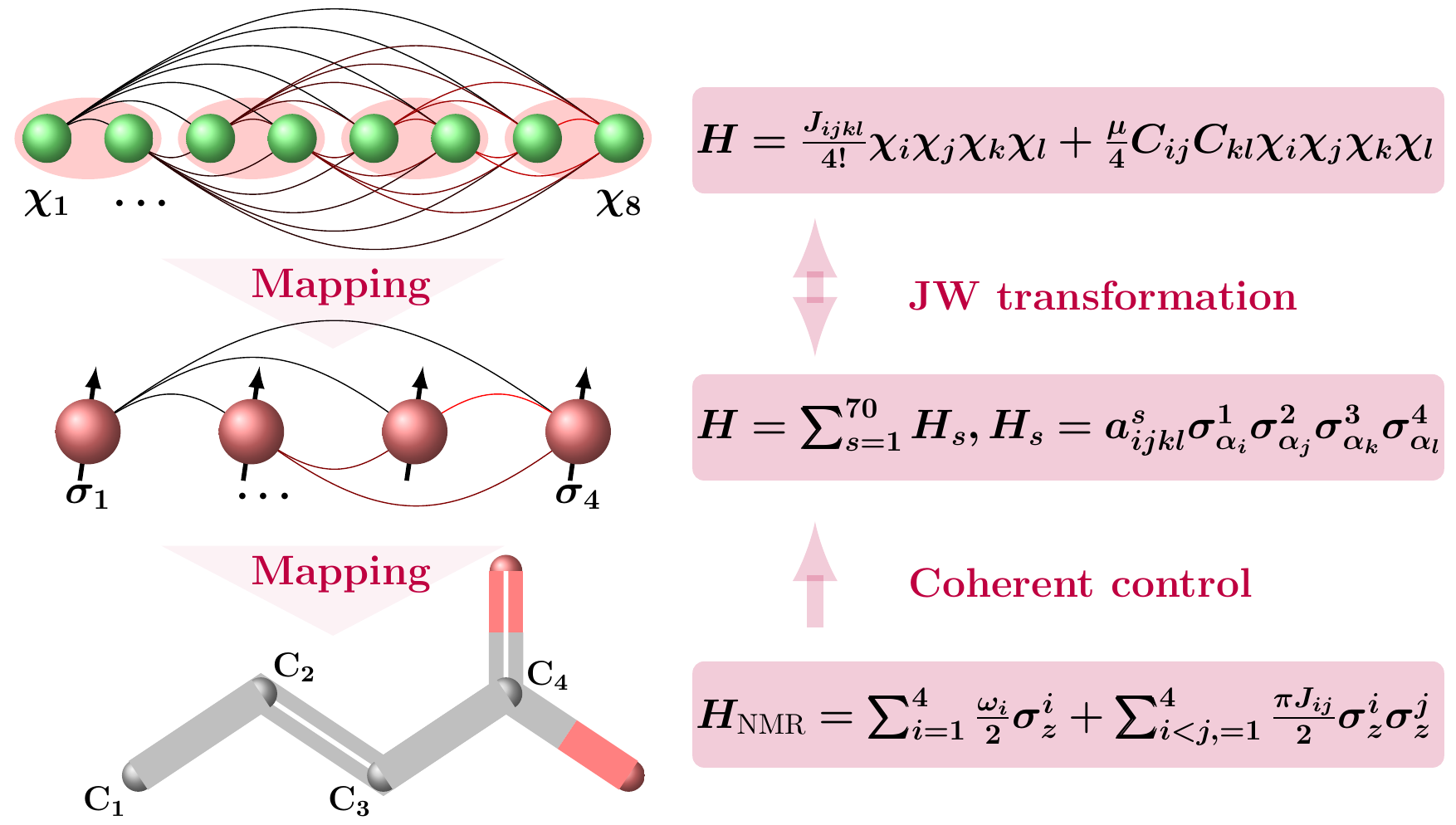}
\caption{Scheme for experimentally simulating the generalized SYK model. The generalized SYK model is firstly mapped onto a spin model via the Jordan-Wigner (JW) transformation, and then we simulate this spin model by the means of  coherent control acting on the physical system of nuclear spins in trans-crotonic acid. The procedure can also be clear from the following change of time evolution operators: $e^{-iH\tau}\leftrightarrow e^{-i\sum_{s=1}^{70}H_s\tau}\approx \left( \prod_{s=1}^{70}e^{-iH_s\tau/n}\right )^n\leftarrow\prod_{j=1}^M e^{-i (H_{\text{NMR}}+H_{\text{C}}(B_j, \phi_j)\tau/M}$. The task in coherent control is to design a pulse sequence for finding the amplitudes $B_j$s and phases $\phi_j$s of radio-frequency fields.}
\label{fig:mapping}
\end{center}
\end{figure}

\textbf{4. Details of experimental simulation.}
The generalized SYK model can be simulated by a fully controllable quantum system, i.e., the four nuclear spins of trans-crotonic acid used in our experiment. As illustrated in Fig. \ref{fig:mapping}, the simulation procedure is stated as follows: After mapping the generalized SYK Hamiltonian (\ref{eq:SYK}) onto a spin model via the Jordan-Wigner transformations, the resulting Hamiltonian (\ref{eq:Hs}) consisting of a sum of 70 local $k$-body ($k\leq 4$) spin interactions, as listed in Table. \ref{tab:spin interaction}, can be effectively simulated by evolving the system forward locally over small, discrete time slices, i.e., simulating the local time evolution operators $e^{-iH_1\tau/n}, e^{-iH_2\tau/n}$, and so on, up to $e^{-iH_{70}\tau/n}$, and repeating $n$ times. Here we use the Trotter-Suzuki approximation decomposition of $e^{-iH\tau}\approx (e^{-iH_1\tau/n}\cdots e^{-iH_{70}\tau/n})^n$, which takes place to within some desired accuracy by choosing sufficiently large $n$, as shown in Fig. 3a. Now the issue of experimental simulation turns into how to implement the discrete time evolutions of local $k$-body spin interactions, $e^{-iH_s\tau/n}=e^{-ia_{ijkl}^s\sigma_{\alpha_i}^1\sigma_{\alpha_j}^2\sigma_{\alpha_k}^3\sigma_{\alpha_l}^4\tau/n}$, for $s=1,\cdots, 70$, by a  NMR quantum simulator. 

The NMR quantum simulator, relying the coherent control of nuclear spins, would allow resolving the above issue of experimentally simulating many-body spin interactions. Let us first review our NMR system in the main paper. Its internal Hamiltonian is $H_{\text{NMR}}=\sum_{i=1}^4\frac{\omega_i}{2}\sigma_z^i+\sum_{i<j, =1}^4\frac{\pi J_{ij}}{2}\sigma_z^i\sigma_z^j$, which consists of $1$-body interactions and $2$-body interactions. The external or control Hamiltonian describing the effect of radio-frequency (RF) pulses is $H_{\text{C}}=\sum_{i=1}^4B_i[\text{cos}(\omega^i_{\text{RF}}t+\phi_i)\sigma_x^i+\text{sin}(\omega^i_{\text{RF}}t+\phi_i)\sigma_y^i]$. By designing a specific pulse sequence (i.e., choosing the appropriate amplitudes $B_i$, frequencies $\omega_{\text{RF}}^i$, phases $\phi_i$ and pulse durations $\tau$), each local time evolution operator $e^{-iH_s\tau/n}$ is readily implemented. For example, the pulse sequences in the rotating frame (we set the reference frequency $\omega_{\text{ref}}=\omega_{\text{RF}}^j=176.053$ MHz in experiments) for simulating 1-, 2-, 3-, and 4-body interactions are given below:
\begin{enumerate}
	\item For $e^{-i\pi J_1 \tau \sigma_x^1/2}$, $[\theta]_x^1$, where $\theta=\pi J_1 \tau$;
	\item For $e^{-i\pi J_{12}\tau\sigma_z^1\sigma_z^2/2}$, $\{\frac{\tau}{4}\} \rightarrow [\pi]_y^4 \rightarrow \{\frac{\tau}{4}\} \rightarrow [\pi]_y^3 \rightarrow \{\frac{\tau}{4}\} \rightarrow [\pi]_y^4 \rightarrow \{\frac{\tau}{4}\}$;
	\item For $e^{-i\pi J_{123}\tau\sigma_z^1\sigma_z^2\sigma_z^3/2}$, $[\frac{\pi}{2}]_x^2 \rightarrow [\pi]_y^2 \rightarrow [\frac{1}{2J_{12}}] \rightarrow [-\frac{\pi}{2}]_{y}^2 \rightarrow  [\frac{J_{123}\tau}{J_{23}}] \rightarrow [-\frac{\pi}{2}]_{y}^2 \rightarrow [\frac{1}{2J_{12}}] \rightarrow [-\frac{\pi}{2}]_{x}^2$;
	\item For $e^{-i\pi J_{1234}\tau\sigma_z^1\sigma_z^2\sigma_z^3\sigma_z^4/2}$, $[\frac{\pi}{2}]_x^2 \rightarrow [\pi]_y^2 \rightarrow [\frac{1}{2J_{12}}] \rightarrow [-\frac{\pi}{2}]_{y}^2 \rightarrow [\frac{\pi}{2}]_x^3 \rightarrow [\pi]_y^3 \rightarrow [\frac{1}{2J_{23}}] \rightarrow [-\frac{\pi}{2}]_{y}^3 \rightarrow [\frac{J_{1234}\tau}{J_{34}}] \rightarrow [-\frac{\pi}{2}]_{y}^3 \rightarrow [\frac{1}{2J_{23}}] \rightarrow [-\frac{\pi}{2}]_{x}^3 \rightarrow [-\frac{\pi}{2}]_{y}^2 \rightarrow [\frac{1}{2J_{12}}] \rightarrow [-\frac{\pi}{2}]_{x}^2$.
\end{enumerate}
Here we denote the above symbols as $[\theta]_{\alpha}^j=e^{-i\theta\sigma_{\alpha}^j}, [\tau_{jk}]=e^{-i\pi J_{jk}\tau\sigma_z^j\sigma_z^k/2}$, and $\{\tau\}=e^{-iH_{\text{NMR}}\tau}$, for simplicity. It could be found that the first case of 1-body interaction can be created by a single pulse of rotation, several refocusing-$\pi$ pulses can realize the specific 2-body interaction, and the cases of 3-,4-body interactions can be implemented by the combination of 1- and 2-body interactions.

\begin{figure}[!htp]
\begin{center}
\includegraphics[width=0.95\linewidth]{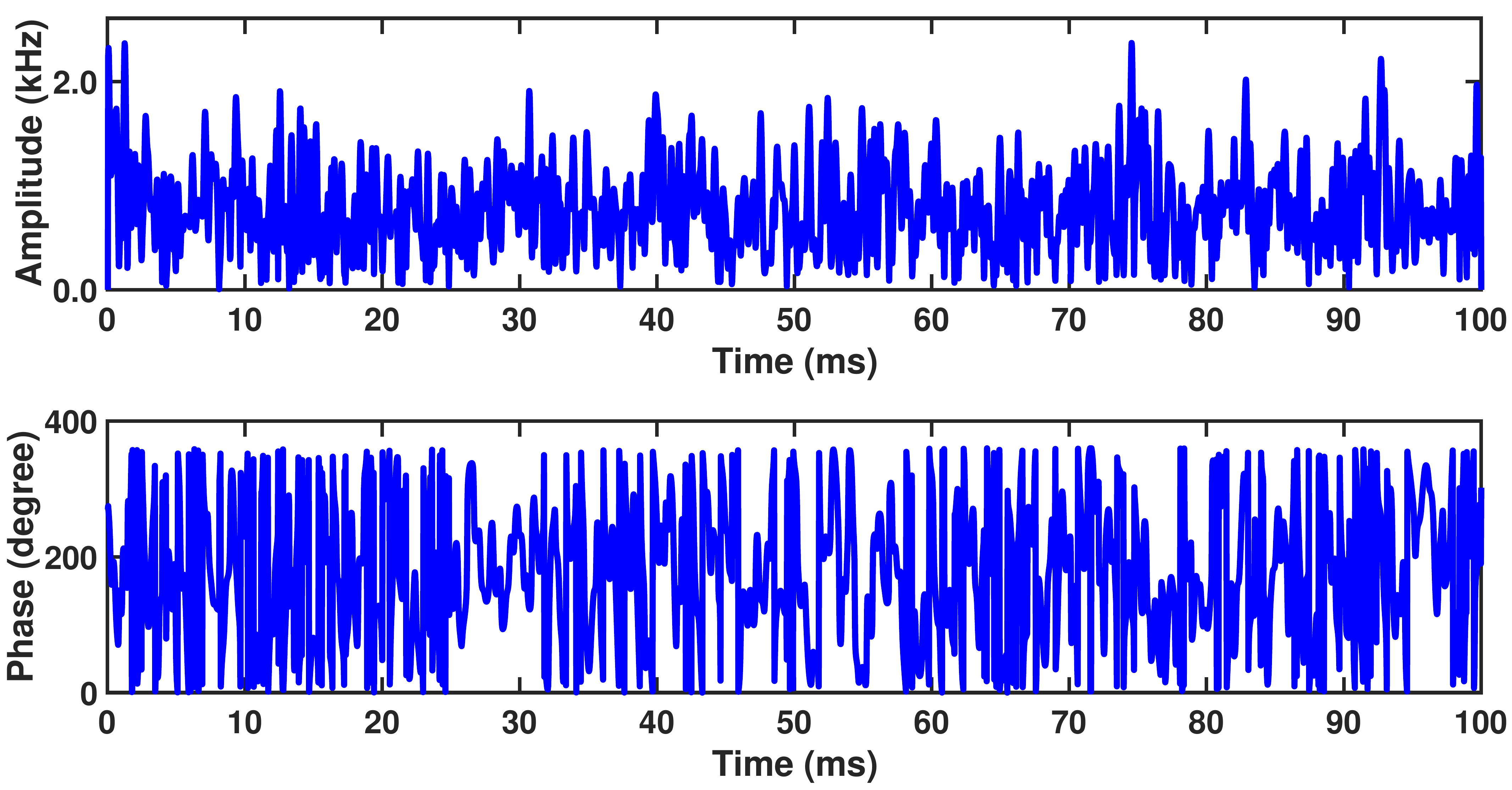}
\caption{The shaped pulse for implementing the evolution of generalized SYK model. Its amplitudes and
phases are shown in the top and in the bottom, respectively. The pulse has 4000 slices, and the total duration is 100 ms.}
\label{fig:waveform}
\end{center}
\end{figure}

For our case of $(e^{-iH_1\tau/n}\cdots e^{-iH_{70}\tau/n})^n=\prod_{j=1}^Me^{-i[H_{\text{NMR}}+H_{\text{C}}(B_j, \phi_j)]\tau/M}$, we employ the gradient ascent pulse engineering (GRAPE) algorithm \cite{Glaser2005} to find its control fields, i.e., the amplitudes $B_j$s and phases $\phi_j$s. The resulting profiles of a shaped pulse with the slices of $M=4000$ and duration of 100 ms  are shown in  Fig. \ref{fig:waveform}. To improve the control performance in simulating the evolution of generalized SYK model, the shaped pulse was designed to have over $99\%$ numerical fidelity in present of $5\%$ the inhomogeneity of radio-frequency fields. 

\begin{figure*}[!htp]
\begin{center}
\includegraphics[width=0.95\linewidth]{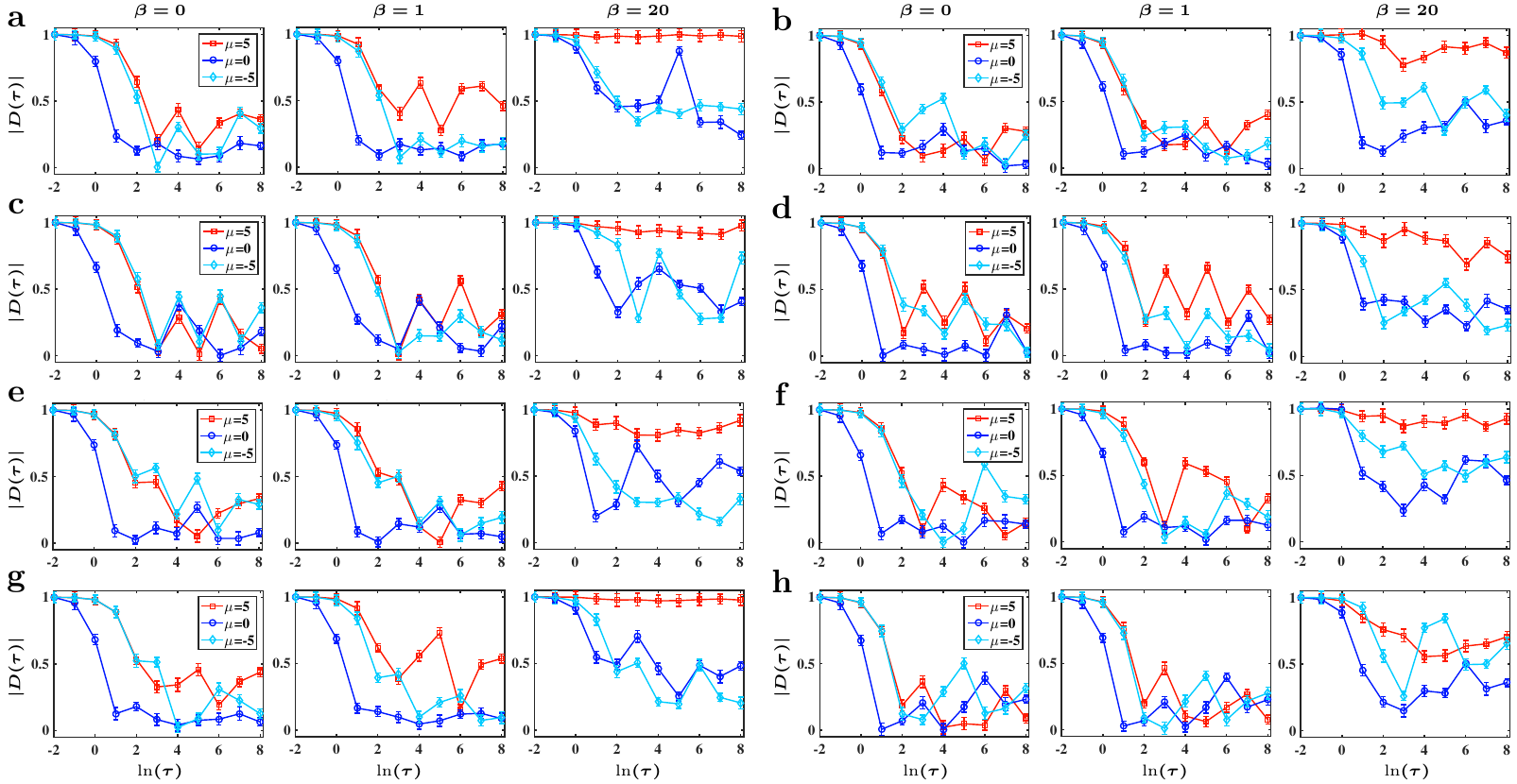}
\caption{The resulting boson correlation functions for $r=1,2,\dots,8$ random samples. The error bars are calculated from the fitting procedure.}
\label{fig:result_sm}
\end{center}
\end{figure*}

\textbf{5. Experimental results for different random samples.} The main experimental results in body paper were obtained by averaging over eight random samples. The boson correlation functions for $r=1,2,\dots,8$ random samples are shown in Figs. \ref{fig:result_sm}(a) $\sim$ \ref{fig:result_sm}(h), respectively.

\begin{figure}[!htp]
\begin{center}
\includegraphics[width=0.7\linewidth]{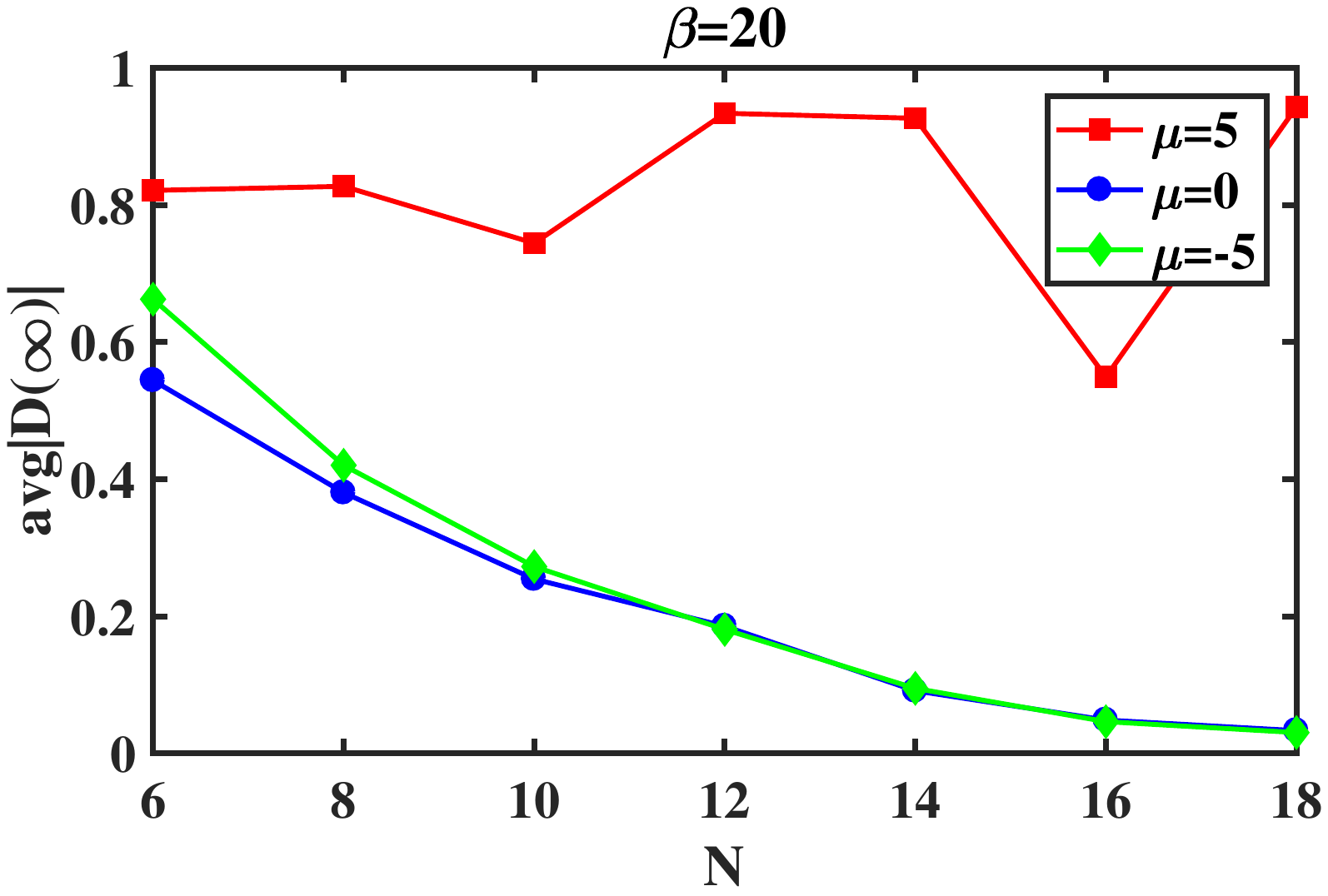}
\caption{System size $N$ dependence of $\text{avg}|D(\infty)|$ at low temperature of $\beta=20$. The boson correlations for $\mu\leq 0$ will decay to zero with the grows of system size; While for $\mu>0$, the boson correlation will saturate to a finite value in thermodynamic limit.}
\label{fig:Dinfinity}
\end{center}
\end{figure}

\textbf{6. Scaling behavior.} It is technically challenging to study a larger system experimentally, thus we resort to numerical simulation to check the scaling behavior. Figure \ref{fig:Dinfinity} shows the system size dependence of $\text{avg}|D(\infty)|$ for $N=6,8,\cdots,18$. In the non-Fermi liquid phase ($\mu\leq0$), the saturate value of boson correlation decays towards zero with system size. In the symmetry breaking phase ($\mu>0$), the saturate value scales towards a finite value in the
thermodynamic limit.

\end{document}